\documentclass[12pt, letterpaper]{article}
\usepackage[titletoc,title]{appendix}
\usepackage{color}
\usepackage{booktabs}
\usepackage[tableposition=top]{caption}

\usepackage[usenames,dvipsnames,svgnames,table]{xcolor}
\definecolor{dark-red}{rgb}{0.75,0.10,0.10} 
\usepackage[margin=1in]{geometry}
\usepackage[linkcolor=dark-red,
            colorlinks=true,
            urlcolor=blue,
            pdfstartview={XYZ null null 1.00},
            pdfpagemode=UseNone,
            citecolor={dark-red},
            pdftitle={Predicting Race and Ethnicity From the Sequence of Characters in a Name}]{hyperref}
\usepackage{geometry} % see geometry.pdf on how to lay out the page. There's lots.
\usepackage{graphicx}                % Handles inclusion of major graphics formats and allows use of 
\usepackage{amsfonts,amssymb,amsbsy}
\usepackage{amsxtra}
\usepackage{natbib}

\usepackage{verbatim}
\setcitestyle{round,semicolon,aysep={},yysep={;}}
\usepackage{setspace}             % Permits line spacing control. Options are \doublespacing, \onehalfspace
\usepackage{sectsty}             % Permits control of section header styles
\usepackage{lscape}
\usepackage{fancyhdr}             % Permits header customization. See header section below.
\usepackage{url}                 % Correctly formats URLs with the \url{} tag
\usepackage{fullpage}             %1-inch margins
\usepackage{multirow}
\usepackage{rotating}
\usepackage{subcaption}

\usepackage{lmodern}
\usepackage{chngcntr}

\title{\Large{Predicting Race and Ethnicity From the\\Sequence of Characters in a Name}\footnote{Data and scripts behind the analysis presented here can be downloaded from \url{http://github.com/appeler/ethnicolr_v2}. This paper is a new version of \citet{sood2018predicting}.}}

\author{Rajashekar Chintalapati\thanks{Rajashekar can be reached at \href{mailto:rajshekar.ch@gmail.com}\texttt{rajshekar.ch@gmail.com}} 
\and Suriyan Laohaprapanon\thanks{Suriyan can be reached at: \href{mailto:suriyant@gmail.com}\texttt{suriyant@gmail.com}}
\and Gaurav Sood\thanks{Gaurav can be reached at \href{mailto:gsood07@gmail.com}\texttt{gsood07@gmail.com}}}

\date{\vspace{.5cm}\normalsize{\today}}

\begin{document}
\maketitle

\begin{abstract}
To answer questions about racial inequality and fairness, we often need a way to infer race and ethnicity from names. One way to infer race and ethnicity from names is by relying on the Census Bureau's list of popular last names. The list, however, suffers from at least three limitations: 1. it only contains last names, 2. it only includes popular last names, and 3. it is updated once every 10 years. To provide better generalization, and higher accuracy when first names are available, we model the relationship between characters in a name and race and ethnicity using various techniques. A model using Long Short-Term Memory works best with out-of-sample accuracy of .85. The best-performing last-name model achieves out-of-sample accuracy of .81. To illustrate the utility of the models, we apply them to campaign finance data to estimate the share of donations made by people of various racial groups, and to news data to estimate the coverage of various races and ethnicities in the news.
\end{abstract}
\clearpage
\doublespace

How often are people of different races and ethnicities covered in the news? What percentage of campaign contributions come from African Americans? Are there racial gaps in healthcare delivery? Do minorities face discrimination in borrowing? To answer such questions, we often need a way to infer race and ethnicity from names. Given the important questions at stake, several researchers have worked on methods to infer race from names \citep[see, e.g.,][]{ambekar2009name, fiscella2006use, imai2016improving, rosenman2022race}. We contribute to this substantial literature.

\section*{Inferring Race and Ethnicity from Names}

\subsection*{Approaches} Researchers have used a variety of approaches to infer race and ethnicity from names. Some researchers have taken advantage of the Census Bureau's list of popular last names \citep[see, e.g.,][]{fiscella2006use}. The approach suffers from three weaknesses---a biased small set of popular last names, a lack of first names, and a decennial update cadence. The bias toward popular names in census data has grave consequences. First, over 82\% of the unique last names included in the census are for NH Whites. This is partly a result of the hard numerical criteria of the last name being shared by 100 or more people which disproportionately affects minorities. Second, last names have a very long tail. Our Florida voter registration database itself has well over a million unique last names. (Part of the long tail may come from errors in how people spell the name but that itself can be seen as its strength as it more realistically generalizes to databases with spelling errors.) The lack of first names is also very consequential. The information in the first name is especially vital for accurately identifying African Americans, whose last names are hard to distinguish from non-Hispanic (NH) whites, and whose first names tend to be distinctive \citep{bertrand2004emily}. 

Other scholars have harvested Wikipedia data on names and their national origins and used crude features of names to build a national origins classifier \citep{ambekar2009name}. The downside of Wikipedia is that the database is small and suffers from a strong bias toward popular people. Yet others have used private communication data and homophily to predict national origin \citep{ye2017nationality}. The limitation here is that the data are private. Lastly, some researchers like us have used voter registration data \citep{sood2018predicting, parasurama2021racebert}.\footnote{There is a parallel literature that combines voter registration data with census data to infer race where we have the name and location of a person \citep[see e.g.,]{imai2016improving, kotovadeep}.} The voter registration data also has its trade-offs. Primarily, the voter registration data which includes the race of the voter is published by a few states, the race variable is often crudely coded, and not everyone is registered to vote \citep{ansolabehere2011gender}. These issues create concerns about its generalizability.

\subsection*{Estimand} We predict the modal race and ethnicity of people with a particular name.\footnote{For models that predict the probability distribution, see \url{https://github.com/appeler/colorquant}}. For last-name models, it means grouping by last name and calculating the modal race and ethnicity of people with that last name. For full-name models, it means calculating the modal race and ethnicity of people with that full name. Hence, the full-name models are estimated on a much finer grain.

\subsection*{Why Model?} If you picked a person at random with the last name Smith from the US in 2010 and were asked to guess this person's race (as measured by the census), the best guess would be based on what is available from the aggregated Census popular last names file. It is the Bayes optimal solution. So what good are last-name predictive models for? Primarily two things. First, the model is useful to predict the race and ethnicity of people whose names are not in the popular last names list. Second, it can be useful in predicting the race and ethnicity of names in databases where names have spelling errors, etc.

\section*{Data}
We exploit Florida Voting Registration data for 2022 \citep{sood_2017}. Florida Voting Registration has information on nearly 15M voters along with their self-reported race. Even though race and ethnicity are self-reported, the limitations of the reporting instrument mean that the quality of the data is debatable. For one, Florida Voter Registration data treats race and ethnicity as one dimension with Hispanic treated as one category. Second, the instrument only allows crude categorization. For instance, Indian Americans are grouped under Asians and Pacific Islanders. In all, we have a self-reported race variable that takes 7 values: Non-Hispanic White, Non-Hispanic Black, Hispanic, Asian or Pacific Islander, Native American, Multi-racial, and Unknown. 

There are very few cases of people not providing their race or ethnicity. We assume that the data are missing at random and remove them from our data. Given that we have very few people who identify as multi-racial and Native American, we condense them into Other. Our final dataset has five categories: Asian/Pacific Islander, Hispanic, Non-Hispanic Blacks, Non-Hispanic Whites, and Other (see Table \ref{table:fl_data}). Lastly, we drop cases where the last name is just 1 character which causes us to lose 75,000 of the roughly 15 million observations. 

\begin{table}[h!]
\centering
\caption{Registered Voters in Florida by Race.}
\begin{tabular}{ l c }
\hline    
Race/Ethnicity & n \\
\hline
NH White  &       9,446,770 \\
Hispanic   &       2,722,579 \\
NH Black  &       2,086,582 \\
Asian      &       329,034 \\
Other      &       424308 \\
\hline
\end{tabular}
\label{table:fl_data}
\end{table}

In addition to the Florida Voter Registration data, we also use the Census Popular Last Name Data \citep{census2010} and the North Carolina Voter Registration Data. 

\section*{Models}
We try models with varying degrees of sophistication starting with the simplest: an edit-distance based KNN model. We follow that by representing names as a 'Bag of Characters' and learning the relationship with ethno-racial categories using Random Forests (RF) and Gradient Boosted (GB) Trees. Lastly, we model the relationship between the sequence of characters and categories using LSTM and Transformer models. For last names for which we have much less data, we also try an LSTM model that leverages synthetic data.

For the models, we process the data in the following way. We first process the names: 1. transform the first and last names to title case, 2. remove any non-alphabetical characters or hyphens, 3. concatenate the last name and first name (ignoring the middle name) for the full name model. Second, we group data by last name or full name and get the conditional means for each of the five ethno-racial categories and find the modal race. These datasets serve as our final data. To learn all the classifiers, we split the corpus into train, validation, and test sets with proportions of .8, .1, and .1 respectively.

\begin{itemize}

    \item \textbf{KNN.} If we had a census of all the names, and the name was the only data we had about a person, then the Bayes Optimal Classifier for predicting the race of the person with that name is the racial category with the largest probability density. As we noted above, this data is unavailable. Assuming we do not have the full list of names or that names in the right table can have spelling errors, the next simplest classifier is an edit-distance-based classifier. We use a bi-char-based distance metric to estimate a KNN classifier. For the last name model, we use cosine distance.\footnote{Experiments suggest that using a more computationally expensive distance metric like Levenshtein distance doesn't improve performance: \url{https://github.com/appeler/edit_names}}. For computational reasons, for the full name model, we use the Jaccard distance via LSH Minhash for 70\% of the training set.

    \item \textbf{RF and GB.} We use a bag of char and bi-char representations of names to learn the relationship between names and ethno-racial categories. We remove infrequent tokens (occurring less than 3 times in the data) and very frequent bi-chars (occurring in over 30\% of the sequences in the data). (See the notebooks for more details.)
    
    \item \textbf{LSTM.} To learn the association between the sequence of characters in names and race and ethnicity, we estimate an LSTM model \citep{graves2005framewise, gers1999learning}. We embed each of the characters in the name in a 256-length real-valued vector before passing it on to two LSTM layers followed by a fully connected layer and a log softmax. We use Negative Log Likelihood Loss and Adam for optimization \citep{kingma2014adam}. (See the notebooks for more details.)

    \item \textbf{Transformer.} We once again start by embedding the characters. We next also estimate the positional encodings before passing on two transformer layers and a log softmax. The other details are the same as in LSTM. (See the notebooks for more details.)
    
    \end{itemize}

\section*{Results}

\subsection*{Last-Name Models}
The best-performing KNN model achieves a 78\% accuracy on the hold-out set (see Table~\ref{table:knn_last_name}. (See \ref{si} for confusion matrices on the hold-out set of all the models.) Of the other models (see Table~ \ref{table:oos_last_name_perf}), only the LSTM model beats the accuracy of the KNN model. The LSTM model is also the top-performing model across subgroups. The model does well when we predict the race and ethnicity of names in the census popular names data as well. The overall accuracy on the census data is about 87\% with accuracy for Non-Hispanic Whites, Hispanics, Asians, and Non-Hispanic Blacks 98\%, 58\%, 42\%, and 26\% respectively. Lastly, we augment the training dataset with synthetic names to see if that improves generalizability. To generate synthetic names, we prompt  ChatGPT to ``give me 10 most common alternate spellings of the last name'' and append the data for popular names. Adding synthetic data provides no appreciable gains in accuracy.

\begin{table}
\centering
\caption{OOS Performance of Various Last Name Models.}
\label{table:oos_last_name_perf}
\begin{tabular}{lccccc}
\toprule
            & RF & GB & LSTM & Transformer & N \\
\midrule
Overall     & 0.55 & 0.75 & 0.81 & 0.73 & 134,898 \\
NH White    & 0.70 & 0.93 & 0.91 & 0.90 & 60,970 \\
NH Black    & 0.19 & 0.12 & 0.50 & 0.09 & 13,726 \\
Hispanic    & 0.66 & 0.80 & 0.84 & 0.79 & 38,961 \\
Asian       & 0.05 & 0.07 & 0.40 & 0.03 & 6,867 \\
Other       & 0.17 & 0.01 & 0.04 & 0.00 & 14,374 \\
\bottomrule
\end{tabular}
\end{table}

Even though the LSTM model based on the Florida Voter Registration data performs well on the census data, a better dataset for modeling patterns in the census may be the census data itself. To that end, we also estimate an LSTM model on the census popular last names data. The model achieves an accuracy of 86\% on the hold-out set. Adding synthetic data improves the accuracy by a percentage point to 87\%.

\subsection*{Full-Name Models}
Next, we discuss the results for the full-name models. The best-performing KNN model achieves an accuracy of 73\%.\footnote{In our experiments on a smaller corpus, we find that the cosine distance method achieves 2--3\% greater accuracy than the approximate LSH Minhash results reported here.} Table~\ref{table:oos_full_name_perf} presents the results for the other full-name models---RF, GB, LSTM, and Transformer. As before, LSTM dominates the other models and is the only model that outperforms KNN. The accuracy of the LSTM model never dips below 63\% for any category except for Other.

\begin{table}
\centering
\caption{OOS Performance of Various Full Name Models.}
\label{table:oos_full_name_perf}
\begin{tabular}{lccccc}
\toprule
            & RF   & GB  & LSTM & Transformer & N \\
\midrule
Overall     & 0.71 & 0.68 & 0.85 & 0.70 & 959,848 \\
NH White    & 0.89 & 0.98 & 0.92 & 0.88 & 573,470 \\
NH Black    & 0.32 & 0.01 & 0.74 & 0.23 & 149,299 \\
Hispanic    & 0.66 & 0.40 & 0.86 & 0.66 & 169,058 \\
Asian       & 0.22 & 0.04 & 0.63 & 0.24 & 27,829 \\
Other       & 0.03 & 0.00 & 0.07 & 0.00 & 40,192 \\
\bottomrule
\end{tabular}
\end{table}

\subsection*{Does Adding the First Name Add Predictive Value?}

As we noted above, our estimand for full-name models is the modal race of a person with a particular full-name. This is a much finer level of inference given the extraordinary uniqueness of first names. But does adding the first name add value? To test that, we compare the performance of the full name model and the last name model on the hold-out set for the full name. The best-performing last name model achieves an accuracy of 74\% compared to 85\% for the best-performing full name model. In particular, note that the accuracy for NH Blacks goes from 21\% to 74\%.

\section*{Applications}
To illustrate the utility of the models we present here, we use them to answer two important questions: 1. Who contributes to politicians? 2. Diversity in the newsroom and the people mentioned. 

To the extent that money buys political influence, it is helpful to examine campaign contributions to politicians. To learn who contributes to politicians, we imputed the race and ethnicity of individual campaign contributors in the 2014 campaign contribution database \citep{bonica2017database} using the Florida full-name LSTM model. As Table~\ref{table:campaign} shows, 89.5\% of the money was contributed by Non-Hispanic Whites, 4.4\% by Non-Hispanic Blacks, 2.6\% by Hispanics, and 3.4\% by Asians. (For comparison, Non-Hispanic Whites were about 64\% of the population in 2014.) 

\begin{table}[ht!]
\centering
\caption{Share of Political Contributions by Race/Ethnicity}
\label{table:campaign}
\begin{tabular}{lc}
  \hline
  Race/Ethnicity & Percentage \\
  \hline
  NH White & 89.5\% \\
  NH Black & 4.4\% \\
  Hispanic & 2.6\% \\
  Asian & 3.4\% \\
  Other & .1\% \\
  \hline
\end{tabular}
\end{table}

To shed light on diversity in the newsroom and people mentioned in the news, we exploited the Top News dataset \citep{willis2023}. The dataset is a collection of articles from news feeds from major news sites like ABC, CBS, CNN, NBC, LA Times, NBC, NYT, Politico, USAT, and WaPo since June 2022. We used the newspaper library \citep{lucas} to parse the data and get the author(s) and the text of each article. We then used NER to extract people's names from the text. As Table~\ref{table:news} shows, Non-Hispanic Whites are overrepresented in the newsroom and in the mentions--78\% of the authors and 73.5\% of the mentions are to Non-Hispanic Whites. On the other hand, African Americans and Hispanics are underrepresented in the newsroom and in mentions.

\begin{table}[ht!]
\centering
\caption{Share of Political Contributions by Race/Ethnicity}
\label{table:news}
\begin{tabular}{lcc}
\hline
Race/Ethnicity & Percentage of Authors & Percentage of Mentions \\
\hline
NH White & 78\% & 73.5\% \\
NH Black & 5.7\% & 9.5\% \\
Hispanic & 7.3\% & 8\% \\
Asian & 8.5\% & 8.6\% \\
Other & 0.4\% & 0.4\% \\
\hline
\end{tabular}
\end{table}

\section*{Discussion}
If you know nothing about a person but the last name, then the best guess for their race and ethnicity is what is provided by the Census popular last name dataset. But as we note, multiple things complicate the picture. Census data is incomplete, census data is biased, the data you may have may contain spelling errors, the data you may have may contain first names, etc. Under those circumstances, using a last-name or full-name model may provide a more accurate prediction. The LSTM models provide excellent accuracy with KNN models surprisingly coming in second.

\clearpage
\bibliographystyle{apsr}
\bibliography{name}
\clearpage

\appendix
\renewcommand{\thesection}{SI \arabic{section}}
\renewcommand\thetable{\thesection.\arabic{table}}  
\renewcommand\thefigure{\thesection.\arabic{figure}}
\counterwithin{figure}{section}
\counterwithin{table}{section}
\section{Supporting Information}\label{si}

\section{Performance of the KNN Models}\label{knn_perf}

\begin{table}[h!]
\centering
\caption{Performance of the KNN (K = 5) Cosine Distance model on the test set.}
\begin{tabular}{lllrr}
\hline
              & precision   & recall   &   f1-score &   support \\
\hline
 asian        & 0.47        & 0.19     &       0.27 &      1,812 \\
 hispanic     & 0.88        & 0.82     &       0.85 &     16,243 \\
 nh\_black     & 0.56        & 0.38     &       0.45 &      5,145 \\
 nh\_white     & 0.77        & 0.90     &       0.83 &     28,183 \\
 other        & 0.22        & 0.04     &       0.07 &      1,450 \\
              &             &          &            &           \\
 accuracy     & -           & -        &       0.78 &     52,833 \\
 macro\_avg    & 0.58        & 0.47     &       0.49 &     52,833 \\
 weighted\_avg & 0.76        & 0.78     &       0.76 &     52,833 \\
\hline
\label{table:knn_last_name}
\end{tabular}
\end{table}

\begin{table}[h!]
\centering
\caption{Performance of the KNN (K = 10) LSH Minhash Jaccard Distance model on the test set.}
\begin{tabular}{lllrr}
\hline
              & precision   & recall   &   f1-score &   support \\
\hline
 asian        & 0.64        & 0.18     &       0.28 &     25,756 \\
 hispanic     & 0.72        & 0.65     &       0.68 &    163,525 \\
 nh\_black     & 0.58        & 0.31     &       0.4  &    133,471 \\
 nh\_white     & 0.75        & 0.91     &       0.82 &    552,737 \\
 other        & 0.30        & 0.02     &       0.03 &     26,373 \\
              &             &          &            &           \\
 accuracy     & -           & -        &       0.73 &    901,862 \\
 macro\_avg    & 0.60        & 0.41     &       0.44 &    901,862 \\
 weighted\_avg & 0.70        & 0.73     &       0.7  &    901,862 \\
\hline
\label{table:knn_full_name}
\end{tabular}
\end{table}

\clearpage
\section{Performance of the Random Forest Models}\label{rf_perf}

\begin{table}[h!]
\centering
\caption{Performance of the Last Name Random Forest model on the test set.}
\begin{tabular}{lllrr}
\hline
              & precision   & recall   &   f1-score &   support \\
\hline
 asian        & 0.62        & 0.17     &       0.26 &      3,648 \\
 hispanic     & 0.86        & 0.83     &       0.85 &     32,461 \\
 nh\_black     & 0.69        & 0.26     &       0.38 &     10,403 \\
 nh\_white     & 0.76        & 0.93     &       0.83 &     56,262 \\
 other        & 0.30        & 0.01     &       0.03 &      2,890 \\
              &             &          &            &           \\
 accuracy     & -           & -        &       0.78 &    105,664 \\
 macro\_avg    & 0.65        & 0.44     &       0.47 &    105,664 \\
 weighted\_avg & 0.76        & 0.78     &       0.75 &    105,664 \\
\hline
\label{table:rf_last_name}
\end{tabular}
\end{table}

\begin{table}[h!]
\centering
\caption{Performance of the Full Name Random Forest model on the test set.}
\begin{tabular}{lllrr}
\hline
              & precision   & recall   &   f1-score &   support \\
\hline
 asian        & 0.79        & 0.19     &       0.31 &     25,756 \\
 hispanic     & 0.84        & 0.72     &       0.77 &    163,525 \\
 nh\_black     & 0.83        & 0.27     &       0.41 &    133,471 \\
 nh\_white     & 0.75        & 0.97     &       0.84 &    552,737 \\
 other        & 0.41        & 0.00     &       0.01 &     26,373 \\
              &             &          &            &           \\
 accuracy     & -           & -        &       0.77 &    901,862 \\
 macro\_avg    & 0.72        & 0.43     &       0.47 &    901,862 \\
 weighted\_avg & 0.77        & 0.77     &       0.73 &    901,862 \\
\hline
\label{table:rf_full_name}
\end{tabular}
\end{table}

\clearpage
\section{Performance of the Gradient Boosting Models}\label{gbm_perf}

\begin{table}[h!]
\centering
\caption{Performance of the Last Name Gradient Boosted Trees model on the test set.}
\begin{tabular}{lllrr}
\hline
              & precision   & recall   &   f1-score &   support \\
\hline
 asian        & 0.67        & 0.07     &       0.13 &      3,648 \\
 hispanic     & 0.83        & 0.80     &       0.81 &     32,461 \\
 nh\_black     & 0.64        & 0.12     &       0.2  &     10,403 \\
 nh\_white     & 0.73        & 0.93     &       0.81 &     56,262 \\
 other        & 0.34        & 0.01     &       0.01 &      2,890 \\
              &             &          &            &           \\
 accuracy     & -           & -        &       0.75 &    105,664 \\
 macro\_avg    & 0.64        & 0.39     &       0.4  &    105,664 \\
 weighted\_avg & 0.74        & 0.75     &       0.71 &    105,664 \\
\hline
\label{table:gb_last_name}
\end{tabular}
\end{table}

\begin{table}[h!]
\centering
\caption{Performance of the Full Name Gradient Boosted Trees model on the test set}
\begin{tabular}{lllrr}
\hline
              & precision   & recall   &   f1-score &   support \\
\hline
 asian        & 0.86        & 0.04     &       0.08 &     25,756 \\
 hispanic     & 0.82        & 0.40     &       0.54 &    163,525 \\
 nh\_black     & 0.77        & 0.01     &       0.03 &    133,471 \\
 nh\_white     & 0.66        & 0.98     &       0.79 &    552,737 \\
 other        & 0.38        & 0.00     &       0    &     26,373 \\
              &             &          &            &           \\
 accuracy     & -           & -        &       0.68 &    901,862 \\
 macro\_avg    & 0.70        & 0.29     &       0.29 &    901,862 \\
 weighted\_avg & 0.70        & 0.68     &       0.59 &    901,862 \\
\hline
\end{tabular}
\label{table:gb_full_name}
\end{table}

\singlespacing
\textsuperscript{*} 
\doublespacing

\clearpage
\section{Performance of the LSTM Models}\label{lstm_perf}

\begin{table}[h!]
\centering
\caption{Performance of the Last Name LSTM model on the test set.}
\begin{tabular}{lllrr}
\hline
              & precision   & recall   &   f1-score &   support \\
\hline
 asian        & 0.55        & 0.41     &       0.47 &      3,647 \\
 hispanic     & 0.89        & 0.85     &       0.87 &     32,445 \\
 nh\_black     & 0.67        & 0.51     &       0.58 &     10,399 \\
 nh\_white     & 0.81        & 0.92     &       0.86 &     56,221 \\
 other        & 0.42        & 0.04     &       0.08 &      2,888 \\
              &             &          &            &           \\
 accuracy     & -           & -        &       0.81 &    105,600 \\
 macro\_avg    & 0.67        & 0.55     &       0.57 &    105,600 \\
 weighted\_avg & 0.80        & 0.81     &       0.8  &    105,600 \\
\hline
\label{table:lstm_last_name}
\end{tabular}
\end{table}

\begin{table}[h!]
\centering
\caption{Performance of the Full Name LSTM model on the test set.}
\begin{tabular}{lllrr}
\hline
              & precision   & recall   &   f1-score &   support \\
\hline
 asian        & 0.65        & 0.63     &       0.64 &     25,752 \\
 hispanic     & 0.85        & 0.86     &       0.85 &    163,505 \\
 nh\_black     & 0.78        & 0.75     &       0.76 &    133,458 \\
 nh\_white     & 0.89        & 0.93     &       0.91 &    552,675 \\
 other        & 0.40        & 0.07     &       0.12 &     26,370 \\
              &             &          &            &           \\
 accuracy     & -           & -        &       0.85 &    901,760 \\
 macro\_avg    & 0.71        & 0.65     &       0.66 &    901,760 \\
 weighted\_avg & 0.84        & 0.85     &       0.84 &    901,760 \\
\hline
\label{table:lstm_full_name}
\end{tabular}
\end{table}

\clearpage
\section{Performance of the Transformer Models}\label{transformer_perf}

\begin{table}[h!]
\centering
\caption{Performance of the Last Name Transformer model on the test set.}
\begin{tabular}{lllrr}
\hline
              & precision   & recall   &   f1-score &   support \\
\hline
 asian        & 0.38        & 0.03     &       0.06 &      3,646 \\
 hispanic     & 0.80        & 0.80     &       0.8  &     32,431 \\
 nh\_black     & 0.48        & 0.09     &       0.16 &     10,398 \\
 nh\_white     & 0.72        & 0.91     &       0.8  &     56,236 \\
 other        & 0.00        & 0.00     &       0    &      2,889 \\
              &             &          &            &           \\
 accuracy     & -           & -        &       0.74 &    105,600 \\
 macro\_avg    & 0.48        & 0.37     &       0.36 &    105,600 \\
 weighted\_avg & 0.69        & 0.74     &       0.69 &    105,600 \\
\hline
\label{table:transformer_last_name}
\end{tabular}
\end{table}

\begin{table}[h!]
\centering
\caption{Performance of the Full Name Transformer model on the test set.}
\begin{tabular}{lllrr}
\hline
              & precision   & recall   &   f1-score &   support \\
\hline
 asian        & 0.56        & 0.25     &       0.34 &     25,753 \\
 hispanic     & 0.67        & 0.67     &       0.67 &    163,509 \\
 nh\_black     & 0.51        & 0.24     &       0.33 &    133,451 \\
 nh\_white     & 0.74        & 0.89     &       0.81 &    55,2674 \\
 other        & 0.28        & 0.00     &       0.01 &     26,373 \\
              &             &          &            &           \\
 accuracy     & -           & -        &       0.71 &    901,760 \\
 macro\_avg    & 0.55        & 0.41     &       0.43 &    901,760 \\
 weighted\_avg & 0.67        & 0.71     &       0.67 &    901,760 \\
\hline
\label{table:transformer_full_name}
\end{tabular}
\end{table}

\end{document}